\documentclass[prl,twocolumn]{revtex4-2}
\usepackage{graphicx}
\usepackage{amssymb}
\usepackage{amsmath}
\usepackage{bm}
\usepackage{color}
\usepackage{physics}

\usepackage[colorlinks,urlcolor=blue,citecolor=blue,linkcolor=blue]{hyperref}
\usepackage{comment}
\usepackage{cleveref}
\usepackage{mathtools}
\usepackage{mathrsfs}

\newcommand{\eb}{\varepsilon_B}
\newcommand{\nn}{\nonumber}
\renewcommand{\k}{\mathbf{k}}
\newcommand{\p}{\mathbf{p}}
\newcommand{\q}{\mathbf{q}}
\newcommand{\Q}{\mathbf{Q}}

\newcommand{\kP}{\mathbf{k}^\prime}

\newcommand{\jfl}[1]{{\color{black}#1}}

\begin{document}

\title{
Enhanced scattering between electrons and exciton-polaritons in a microcavity}

\author{Guangyao Li}
\affiliation{School of Physics and Astronomy, Monash University, Victoria 3800, Australia}
\affiliation{ARC Centre of Excellence in Future Low-Energy Electronics Technologies, Monash University, Victoria 3800, Australia}

\author{Olivier Bleu}
\affiliation{School of Physics and Astronomy, Monash University, Victoria 3800, Australia}
\affiliation{ARC Centre of Excellence in Future Low-Energy Electronics Technologies, Monash University, Victoria 3800, Australia}

\author{Meera M. Parish}
\affiliation{School of Physics and Astronomy, Monash University, Victoria 3800, Australia}
\affiliation{ARC Centre of Excellence in Future Low-Energy Electronics Technologies, Monash University, Victoria 3800, Australia}

\author{Jesper Levinsen}
\affiliation{School of Physics and Astronomy, Monash University, Victoria 3800, Australia}
\affiliation{ARC Centre of Excellence in Future Low-Energy Electronics Technologies, Monash University, Victoria 3800, Australia}

\begin{abstract}
The interplay between strong light-matter interactions and charge doping 
represents an important frontier in the pursuit of exotic many-body physics and optoelectronics. Here, we consider a simplified model of a two-dimensional semiconductor embedded in a microcavity, where the
interactions between electrons and holes are strongly screened, allowing us to develop a diagrammatic formalism for this system 
with an analytic expression for the exciton-polariton propagator. We apply this to the scattering of spin-polarized polaritons and electrons, and show that this is strongly enhanced compared with exciton-electron interactions. As we argue, this counter-intuitive result is a consequence of the shift of the collision energy due to the strong light-matter coupling, and hence this is a generic feature that applies also for more realistic electron-hole and electron-electron interactions. We furthermore demonstrate that the lack of Galilean invariance inherent in the light-matter coupled system can lead to a narrow resonance-like feature for polariton-electron interactions
close to the polariton inflection point. 
Our results are potentially important for realizing tunable light-mediated interactions between charged particles.  
\end{abstract}

\maketitle

Exciton-polaritons are hybrid light-matter quasiparticles resulting from the strong coupling between 
excitons (bound electron-hole pairs in a semiconductor)  
and microcavity photons~\cite{DengRevMod10,CiutiRevMod13,ByrnesNatPhys14}.
Their photonic component endows them with an exceptionally small mass and the potential for 
optical control, which is necessary 
for realizing 
coherent phenomena such as polariton lasing or
Bose-Einstein condensation~\cite{Deng2003,BEC06,BaliliScience07,UtsunomiyaNatPhys08,RoumposNatPhys11} at elevated temperatures and in a variety of tailored geometries \cite{Cerda2010,SalaPRX15,GaoTalbot16,Ohadi2017,WhittakerPRL18}.
Their excitonic component furthermore 
gives polaritons the ability to interact pairwise among themselves and with other particles, which can give rise to 
polariton superfluidity \cite{Amo2009,Sanvitto2010,KohnlePRL11} and 
photon quantum correlations~\cite{Munoz2019,Delteil2019}.
Of particular interest is the interplay between polaritons and 
electrons,
a scenario which can be achieved via photoexcitation or doping of a two-dimensional (2D) semiconductor embedded in a microcavity~\cite{GabbayPRL07,SmolkaScience14,Baeten2014,SidlerNatPhys16,Chervy2020}.
Most recently, the atomically thin transition metal dichalocogenides (TMDs) have emerged as promising platforms for this purpose~\cite{Ardizzone2019}. 
The combined polariton-electron system can lead to enhanced polariton-polariton interactions~\cite{KnuppelNature2019,Tan2020,Emmanuele_2020}, 
as well as the possibility of optically engineered electronic phases such as polariton-mediated superconductivity~\cite{LaussyPRL10,CotlePRB16,KavokinNatMater2016}.

To investigate and potentially exploit the properties of    
electron-rich polariton systems, a necessary ingredient is a microscopic description 
of polariton-electron interactions. 
However, this is challenging to achieve theoretically since one must solve a multi-body problem that involves at least three charged particles as well as a photonic component. 
Indeed, the case of electron-\textit{exciton} scattering has only very recently been studied theoretically with exact state-of-the-art techniques~\cite{CombescotPRX2017,Fey2020}. To date, studies of polariton-electron scattering have primarily involved calculations within the lowest-order Born approximation~\cite{RamonPRB02,Malpuech2002,SnokePRB2010,Tartakovskii2003,Shahnazaryan2017}, where the excitonic part is treated as an inert object, unaffected by the coupling to photons.
While recent studies treat polariton-electron interactions beyond the Born approximation~\cite{SidlerNatPhys16,Tan2020}, these only consider polaritons at zero momentum and they ignore the composite nature of the excitonic component.
Thus, a complete description of polariton-electron interactions 
is still lacking.

In this Letter, we solve the polariton-electron problem 
for the case where the spins are polarized and 
the interactions between charges are strongly screened, 
such that they correspond to short-range contact interactions. 
This simplification for the polariton system has been widely used in the literature~\cite{Yamaguchi2012,Hanai2017,Hu2019QuantumFI}
and allows us to obtain an analytic expression for the polariton propagator that captures the non-perturbative effects of the light-matter coupling on the excitonic part~\cite{Khurgin2001,JesperPRR19}.
We then obtain the polariton-electron interaction $T$ matrix using a three-body diagrammatic approach~\cite{Skorniakov1957,BedaquePRC1998,BrodskyPRA06,LevinsenPRA06,Levinsen2011,Ngampruetikorn2013} that has been successfully applied to cold-atom experiments and neutron scattering~\cite{Varenna2008,PetrovLesHouches2010,
HammerRMP13}.
We find that the strength of polariton-electron scattering is strongly enhanced compared to the exciton-electron case, which is 
the opposite of what is expected based on the standard Born approximation. In particular, we reveal a resonance-like enhancement of elastic scattering at finite polariton momentum, which is intimately connected to the non-Galilean nature of the polariton system. We argue that these are generic results that should also hold for more realistic
interactions in the semiconductor, such as Coulomb interactions.

\paragraph{Model.---} We consider a spin-polarized 2D semiconductor in a planar microcavity. The system is described by the Hamiltonian (we set $\hbar$ and the system area to 1)
\begin{align}\label{eq:Hamiltonian}
\hat{H}=&\sum_\k \left(\epsilon_\k e^\dagger_\k e_\k+\epsilon_\k h^\dagger_\k h_\k \right)
 - V_0 \sum_{\k \k'\q} e^\dagger_{\k}h^\dagger_{\q-\k}h_{\q-\kP}e_{\kP} \nn \\
&\hspace{-1mm}
+\sum_\k(\omega+\epsilon^c_{\k})c^\dagger_{\k}c_{\k}+g\sum_{\k\,\q}\left( e^\dagger_{\k} h^\dagger_{\q-\k} c_{\q} +
\rm{H.c.} 
\right).
\end{align}
Here, $e^\dag_\k$, $h^\dag_\k$, and $c^\dag_\k$ create an electron, hole, or photon, respectively, with momentum $\k$. The top line describes electrons and holes which, for simplicity, we take to have contact interactions of strength $V_0$. 
Owing to Pauli exclusion, the  electrons (and holes) do not interact among themselves. 
For simplicity, we take these to have the same mass $m$ (as is approximately the case in TMDs~\cite{Korm_nyos_2015,TMDREV2018}), such that their dispersion is $\epsilon_\k=|\k|^2/2 m \equiv k^2/2 m$. The second line describes the cavity photons with dispersion $\epsilon_\k^c=k^2/2 m_{c}$ and mass $m_c=10^{-4}m$ \cite{DengRevMod10}, where $\omega$ is the zero-momentum resonant frequency in the absence of the active medium. All energies are defined with respect to the semiconductor band gap. 
The last term corresponds to the light-matter coupling with strength $g$ (we applied the rotating wave approximation). Both the electron-hole and light-matter interactions \jfl{are functions of}
an ultraviolet cutoff $\Lambda$ on the relative electron-hole momentum.
\jfl{In the following, we develop a renormalized low-energy theory where we take $\Lambda \to \infty$ and we relate the bare interaction parameters appearing in Eq.~\eqref{eq:Hamiltonian} to physical observables.}

\begin{figure}
	\centering
	\includegraphics[width=\linewidth]{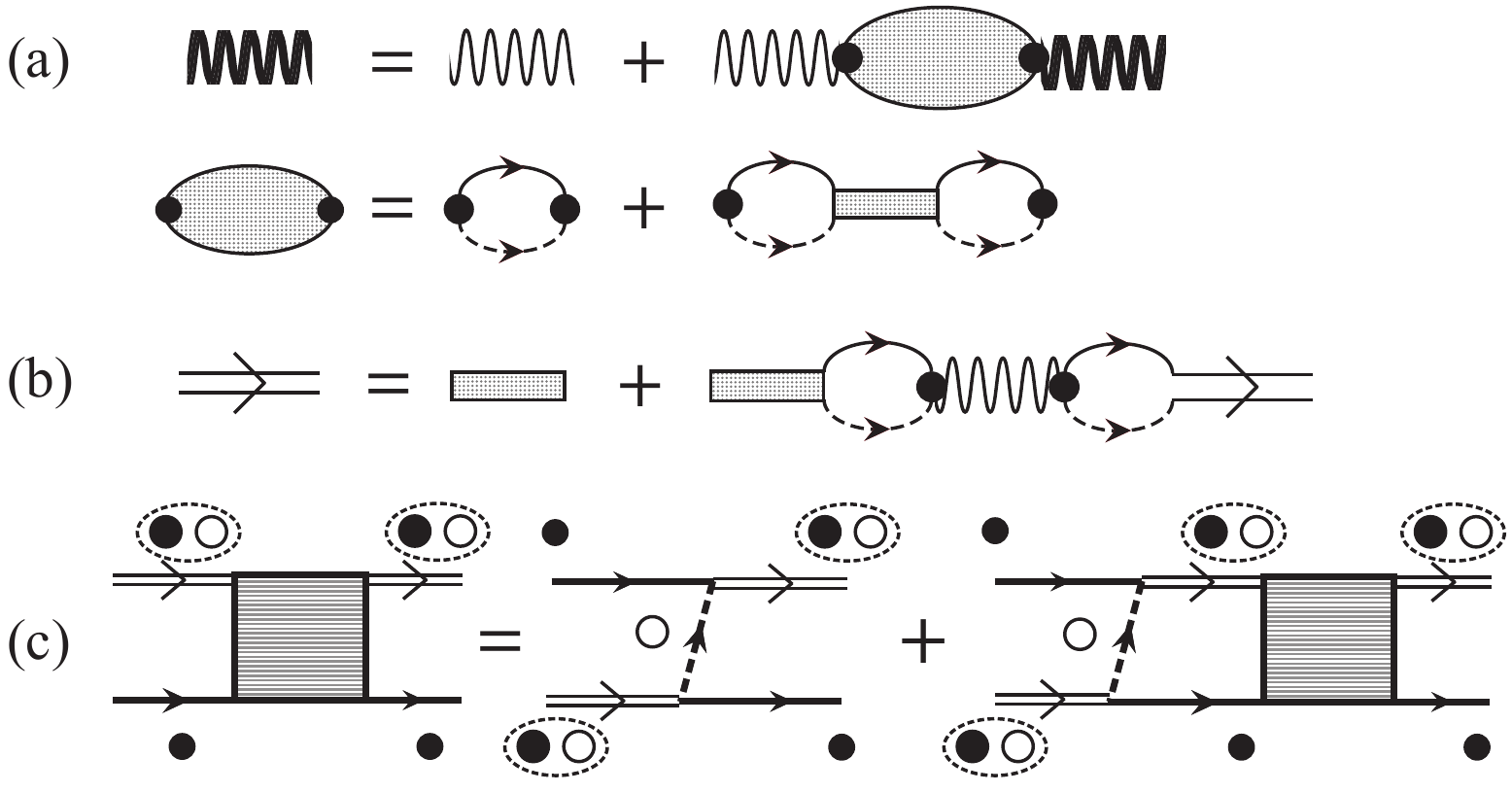}
	\caption{(a) Dyson equation for the dressed photon propagator (thick wavy line) in terms of the bare propagator (thin wavy line) and the self energy (shaded ellipse). The self energy consists of the two terms in Eq.~\eqref{eq:Sigma}, where the thin lines with arrows are fermion propagators and the shaded rectangle is the electron-hole $T$ matrix defined in Eq.~\eqref{eq:Tmat0}. Black dots represent the light-matter coupling $g$. (b) Polariton propagator (double lines with an arrow) given by repeated interactions between excitons, dressed photons, and electron-hole pairs.  (c) Polariton-electron scattering, as encoded in the polariton-electron $T$ matrix (shaded square). Black, white, and dashed circles represent electrons, holes, and polaritons, respectively.}
	\label{fig:diagrams}
\end{figure}

\paragraph{Photon and polariton propagators.---}
Inside the semiconductor 
microcavity, photons are modified by repeated interactions with 2D electron-hole pairs, and the resulting dressed photon is characterized by the self-energy $\Sigma$~\cite{FetterBook}. 
As illustrated in Fig.~\ref{fig:diagrams}(a), this leads to the Dyson equation for the 
dressed photon propagator at momentum $\Q$ and energy $E$~\cite{JesperPRR19,ePlong}
\begin{equation}\label{eq:Dyson}
D(\Q,E) 
=\frac{1}{D ^{-1}_0(\Q,E)-\Sigma(\Q,E)},
\end{equation}
where the bare propagator $D_0(\Q,E)=1/(E-\omega-\epsilon_{\Q}^c)$
has poles that coincide with the resonant cavity photon dispersion.
Here, and in the following, we assume that the energy poles are shifted slightly into the lower half of the complex plane, i.e., we have retarded propagators~\cite{FetterBook}.

The photon self-energy in Fig.~\ref{fig:diagrams}(a) is composed of two terms: $\Sigma(\Q,E)=\Sigma^{(1)}(E-\epsilon_\Q/2)+\Sigma^{(2)}(E-\epsilon_\Q/2)$~\cite{JesperPRR19}. 
These contain all possible processes that involve the excitation of an electron-hole pair, and they thus only depend on the energy in the electron-hole center-of-mass frame. Within the model~\eqref{eq:Hamiltonian}, we have
\begin{align}\label{eq:Sigma}
\Sigma^{(1)}(E)&=g^2\,\Pi(E),\quad 
\Sigma^{(2)}(E)=g^2\, \Pi^2(E)\, {\cal T}_0(E).
\end{align}
Here, $\Pi(E)\equiv \sum_{\k}^\Lambda \frac{1}{E-2\epsilon_{\k}}$ is the electron-hole pair bubble and ${\cal T}_0$ is the electron-hole $T$ matrix (see, e.g., Ref.~\cite{LevinsenBook15})
\begin{align} \label{eq:Tmat0}
{\cal T}_0(\Q,E)\equiv {\cal T}_0(E-\epsilon_\Q/2)=\frac{4\pi/m}{-\ln[(E-\epsilon_\Q/2)/\eb]+i\pi},
\end{align}
where $\eb\equiv 1/ma_X^2$ is the 1$s$ exciton binding energy, with corresponding Bohr radius $a_X$~\footnote{\jfl{Note that the renormalization process relates the bare electronic interaction to the exciton binding energy via $1/V_0=-\Pi(-\eb)$~\cite{ePlong}}.}. We see that the pair bubble $\Pi$ diverges logarithmically when the momentum cutoff $\Lambda$ is taken to infinity, which implies that $\Sigma^{(2)}/\Sigma^{(1)} = \Pi {\cal T}_0 \to \infty$.
Hence, to obtain a finite  
coupling between light and matter in this limit, 
we require $g \sim 1/\ln\Lambda$ and therefore $\Sigma^{(1)}\to0$~\cite{ePlong}.

We now wish to relate the parameters $g$ and $\omega$ to real observables in experiment. Since the poles of the photon propagator $D$ 
correspond to the polariton spectrum, we can compare them 
to the quasiparticle energies obtained from the coupled-oscillator model~\cite{HopfieldPR1958,DengRevMod10,CiutiRevMod13} which is typically used to fit experimental data.
Here, one assumes that $\eb$ 
is larger than all other relevant energy scales, such that the exciton's composite nature can be neglected. 
This gives the quasiparticle dispersions
\begin{align} \nn
  E^{\rm osc}_\pm(\Q)= \frac{\frac{\epsilon_\Q}2+\delta+\epsilon_\Q^c\pm \sqrt{(\frac{\epsilon_\Q}2-\delta-\epsilon_\Q^c)^2+4\Omega^2}}2 -\eb,
\end{align}
with $-$ ($+$) referring to the lower (upper) polariton. The relevant physical parameters are the photon-exciton detuning $\delta$ and the Rabi coupling $\Omega$, 
which we identify in our theory by performing a perturbative analysis
of Eq.~\eqref{eq:Dyson} close to the exciton energy $-\eb$
(for additional details, see Ref.~\cite{ePlong}). This yields:
\begin{align}
\delta=\omega+\eb-\Omega^2/(2\eb),\qquad
\Omega=-g\sqrt{Z_X}\Pi(-\eb),
\end{align}
where $Z_X\equiv4\pi \eb/m$ is the residue of ${\cal T}_0$ at the exciton pole.
This procedure finally leads to the dressed photon propagator
\begin{align}
  D(\Q,E) & =\frac1{D_0^{-1}(\Q,E)-\frac{\Omega^2}{Z_X}{\cal T}_0(\Q,E)}, \label{eq:dressedphoton}
\end{align}
which is now fully expressed in terms of experimentally measurable quantities. We note that Ref.~\cite{Hu2019QuantumFI} recently presented an alternative renormalization procedure for the contact interaction potential; however, our method has the advantage that it is fully analytic and it does not involve any infrared divergence.

One particularly important parameter is the polariton photon fraction $|C_\pm(\Q)|^2$, otherwise known as a Hopfield coefficient. In our present formulation, this is the residue of the propagator at the polariton energy 
$E_\pm(\Q)$ (which is determined numerically from the propagator's poles). Expanding Eq.~\eqref{eq:dressedphoton} at the pole yields
\begin{align}
  |C_\pm(\Q)|^2=\left(1+\frac{Z_X/Z_\pm(\Q)}{\Omega^2D_0^2(\Q,E_\pm(\Q))}\right)^{-1}.
  \label{eq:photonfraction}
\end{align}
Here, $Z_\pm(\Q)\equiv4\pi|E_\pm(\Q)-\epsilon_\Q/2|/m$ is the polariton generalization of $Z_X$. For the lower polariton ($-$), this is typically well approximated by $Z_X$, in which case Eq.~\eqref{eq:photonfraction} exactly matches the expression obtained for two coupled oscillators~\cite{HopfieldPR1958,DengRevMod10,CiutiRevMod13,ePlong}.

We now \textit{define} the polariton propagator as consisting of all interaction processes between an electron and a hole, as illustrated in Fig.~\ref{fig:diagrams}(b). This is a natural definition for the purposes of calculating polariton-electron scattering, since the pairwise interactions only involve the electronic degrees of freedom, not the photons~\footnote{\jfl{Note that processes in electron-hole scattering that either start or end with a photon line renormalize to zero when the cutoff $\Lambda\to\infty$~\cite{ePlong}.}}. 
Therefore, the polariton propagator is simply a dressed electron-hole $T$ matrix
\begin{align}
  {\cal T}(\Q,E)
  & = \frac1{{\cal T}_0^{-1}(\Q,E)-\frac{\Omega^2}{Z_X}D_0(\Q,E)}.
\end{align}
Note the similarity to Eq.~\eqref{eq:dressedphoton}.
The polariton exciton fraction $|X_\pm(\Q)|^2 \equiv 1-|C_\pm(\Q)|^2$ is related to the residue at the pole $E_\pm(\Q)$,
which is given by $Z_\pm(\Q) |X_\pm(\Q)|^2$. 

\paragraph{Polariton-electron interactions.---} 
We now calculate the interaction strength between an electron and a lower polariton, as illustrated in Fig.~\ref{fig:diagrams}(c).
This follows the diagrammatic approach 
first developed 
in the context of neutron-deuteron scattering~\cite{Skorniakov1957,BedaquePRC1998}, and later applied to cold atomic gases~\cite{BrodskyPRA06,LevinsenPRA06,Levinsen2011,Ngampruetikorn2013}.
The diagrams in Fig.~\ref{fig:diagrams}(c) resemble those of 
exciton-electron scattering~\cite{ePlong}, but there are two major, albeit hidden, differences. First, the very nature of the polariton as a quasiparticle formed from components of different masses means that Galilean invariance is broken. 
Second, since the electron interacts only with the excitonic part of the polariton, 
the scattering of electrons and polaritons can be viewed as strongly \textit{off-shell} exciton-electron interactions, where the collision energy is shifted by the light-matter coupling~\cite{Bleu2020}. These unusual characteristics lead to strong qualitative differences from the conventional Born approximation treatment of polariton-electron scattering~\cite{RamonPRB02,Malpuech2002,SnokePRB2010}, as we shall demonstrate in the following.

The 
polariton-electron $T$ matrix in Fig.~\ref{fig:diagrams}(c) satisfies the integral equation~\cite{ePlong}
\begin{align}
&\hspace{-2mm}T_{eP}(\p_1,\p_2)=-\frac{|X_-(\p_2)|^2  Z_-(\p_2)}{E-\epsilon_{\p_1}-\epsilon_{\p_2}-\epsilon_{\p_1+\p_2}}
\nn
\\
&\hspace{-2mm}\,-\sum_\q \frac{1}{E-\epsilon_{\p_1}-\epsilon_{\q}-\epsilon_{\p_1+\q}}
{\cal T} (\q,E-\epsilon_{\q}) T_{eP} (\q,\p_2).
\label{eq:ePtMatrix}
\end{align}
Here, the electron [polariton] is taken to have momentum $\p_i$ [$-\p_i$] and energy $\epsilon_{\p_i}$ [$E-\epsilon_{\p_i}$], respectively, where $i=1,2$ corresponds to incoming or outgoing particles. 
We consider zero center-of-mass momentum, so that angular momentum is a good quantum number, but our results should also apply to finite center-of-mass momentum since they are relatively insensitive to the electron momentum.
Both minus signs on the right-hand-side of Eq.~\eqref{eq:ePtMatrix} originate from the exchange of identical electrons.

In the following, we consider elastic scattering where $E=E_-(\p_2)+\epsilon_{\p_2}$ is the total energy of the electron-lower-polariton system and $|\p_1|=|\p_2| \equiv Q$, where we note that Eq.~\eqref{eq:ePtMatrix} must be solved as an integral equation in terms of the incoming momentum and thus the latter condition should be taken at the end of the calculation~\cite{ePlong}. Since angular momentum is conserved, 
we remove all angular dependence from the problem by projecting Eq.~\eqref{eq:ePtMatrix} onto the $s$-wave channel~\cite{Ngampruetikorn2013}, 
which is the dominant channel when $Q a_X \ll 1$. 
With these manipulations,  
we arrive at the normalized $s$-wave polariton-electron interaction $T$ matrix, $T_{eP}(Q)\equiv 
\left<T_{eP}(\p_1,\p_2)\right>_\theta$. 

\begin{figure}
	\centering
	\includegraphics[width=1\linewidth]{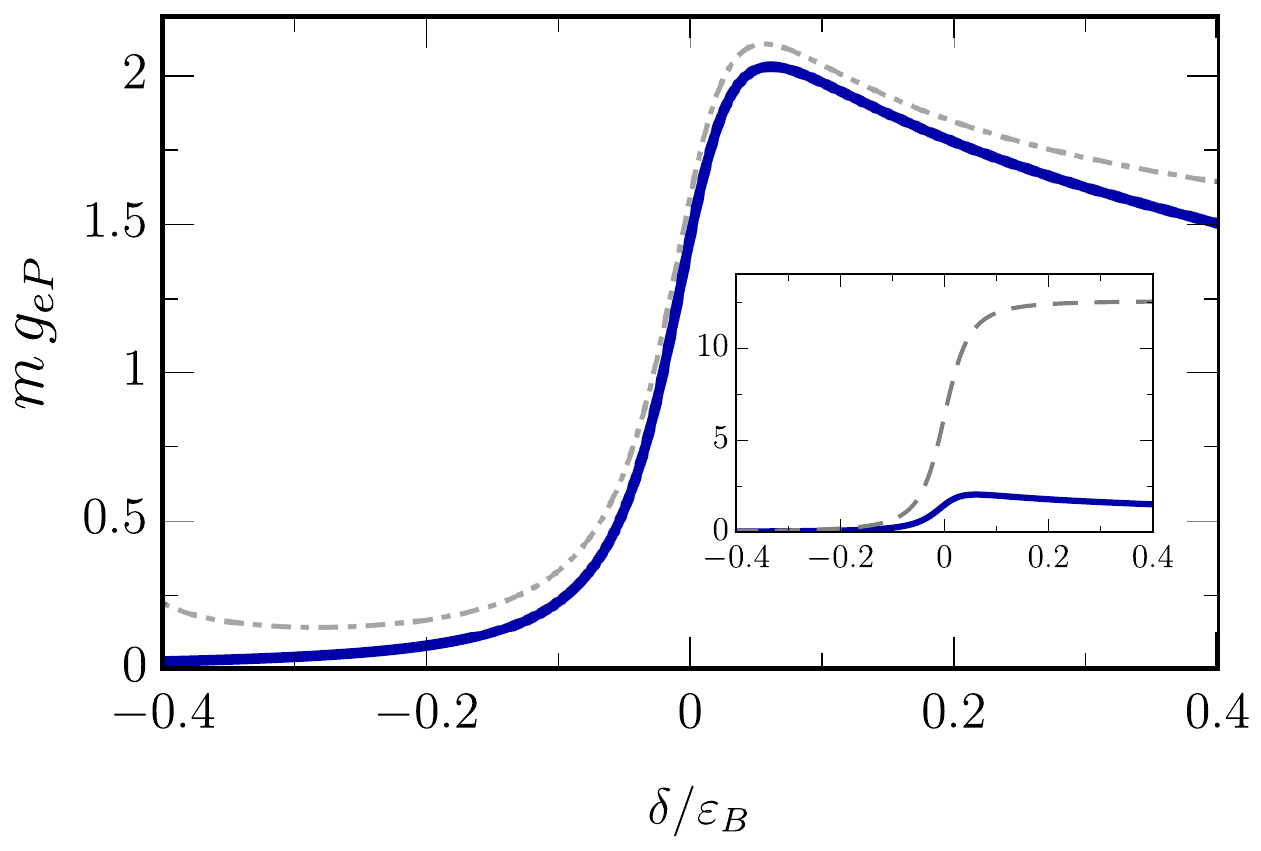}
	\caption{Polariton-electron interaction strength as a function of photon-exciton detuning, \jfl{taking $\Omega/\eb=0.025$ as in a MoSe$_2$, MoS$_2$, or WSe$_2$ monolayer}. 
	The exact diagrammatic calculation (blue solid line) compares well with the off-shell exciton-electron scattering approximation in Eq.~\eqref{eq:Tolistyle} (gray dash-dotted line). Inset: the Born approximation (gray dashed line) deviates from the exact calculation in a large parameter region.}
	\label{fig:Tepzero}
\end{figure}

\paragraph{Scattering of slow particles.---} As a first illustration, we consider scattering in the limit of zero momentum. 
The $T$ matrix $T_{eP}(0)$ is then precisely the polariton-electron coupling constant $g_{eP}$, 
a parameter that serves as an input in 
many-body theories of polaritons and electrons~\cite{Rapaport2000,kyriienko2019nonlinear}. Our results are shown in Fig.~\ref{fig:Tepzero} for the case of $\Omega/\eb=0.025$ relevant for a MoSe$_2$~\cite{dufferwiel2015exciton}, MoS$_2$~\cite{liu2015strong}, or WSe$_2$~\cite{lundt2016room,He2014} monolayer (for WS$_2$, $\Omega/\eb\simeq0.05$~\cite{Flatten2016}).
We see that the interaction strength quickly increases with increasing exciton fraction $|X_-(0)|^2$ from negative to zero detuning, while it has a peak at small positive detuning. By contrast, we find that the result from the Born approximation, $g_{eP}^{\rm Born}=
\jfl{4\pi \eb a_X^2}|X_-(0)|^2$ --- obtained by neglecting the second line in Eq.~\eqref{eq:ePtMatrix} --- is monotonic and greatly overestimates the interaction strength. Indeed, one can show that the Born approximation provides an \textit{upper bound} on the interaction strength in the absence of a trion bound state~\cite{ePlong}.

Within our exact calculation, the behavior of $g_{eP}$ is dominated by the strong light-matter coupling, which determines the collision energy of the scattering processes in the matter component.
To demonstrate this idea of off-shell exciton-electron scattering, we compare our results with that obtained from the universal behavior of low-energy exciton-electron scattering:
\begin{align}
    g_{eP} \simeq |X_-(0)|^2 \frac{3\pi/m}{-\ln\left[-\frac{E_-(0)+\eb}{\varepsilon_{1}}\right]}.
    \label{eq:Tolistyle}
\end{align}
In the present context, low energy implies $\Omega,|\delta|\ll \eb$ which is typically a good approximation in the TMDs. 
In the case of contact electron-hole interactions, the energy scale $\varepsilon_1=\frac{3}{4ma_{eX}^2}$, with exciton-electron scattering length $a_{eX}\simeq 1.26a_X$~\cite{Ngampruetikorn2013}. As seen in Fig.~\ref{fig:Tepzero}, this approximation works extremely well across a range of detunings 
provided $|\delta|$ is not too large. The energy dependence of Eq.~\eqref{eq:Tolistyle} also explains why $g_{eP}$ eventually decreases with increasing positive detuning in Fig.~\ref{fig:Tepzero}. 
Note that Eq.~\eqref{eq:Tolistyle} predicts a spurious resonance at large negative detuning, i.e., outside its regime of validity. 

We emphasize that Eq.~\eqref{eq:Tolistyle} represents the universal form of low-energy polariton-electron scattering, and it therefore also applies to exciton-polaritons in GaAs quantum wells when $\Omega\ll\eb$, or for more realistic electronic interactions in the TMDs~\cite{Keldysh1979}. In each case, \jfl{the low-energy behavior is guaranteed by the long-range $-1/r^4$ form of the interaction between a charge (electron) and a dipole (exciton) at separation $r$~\cite{landau2013quantum}, while} $\varepsilon_1\sim\eb$ should be obtained from first-principles calculations \jfl{such as those recently carried out in Ref.~\cite{Fey2020} for the TMDs. However, we emphasize that since $\varepsilon_1$ appears under a logarithm, we may expect Eq.~\eqref{eq:Tolistyle} to be rather insensitive to the precise value of $\varepsilon_1/\eb$, and consequently that it remains \textit{quantitatively} accurate beyond the strongly screened approximation used here.}

The fact that the polariton-electron interactions are non-vanishing is a dramatic consequence of broken Galilean invariance in the polariton system. According to two-dimensional scattering theory~\cite{landau2013quantum,AdhikariAJP86}, the polariton-electron interaction should in fact approach zero as $\sim1/(m_C\ln[-1/Qa_X])$ when $Q\to0$~\cite{ePlong}. However, this logarithmic term only becomes relevant when it is comparable to $\sim 1/m$, which translates into $Q\sim \exp(-10^4)/a_X$. This momentum scale is never relevant in any experiment since it requires a system that is much larger than 
the size of the known universe. 
By contrast, the relevant momentum scale below which exciton-electron interactions vanish logarithmically is only $1/a_X$.  
Thus, our results demonstrate that polariton-electron interactions are typically \textit{enhanced} compared to the exciton-electron case, even though the polariton contains a non-interacting photonic component.

\begin{figure}
\centering
	\includegraphics[width=1\linewidth]{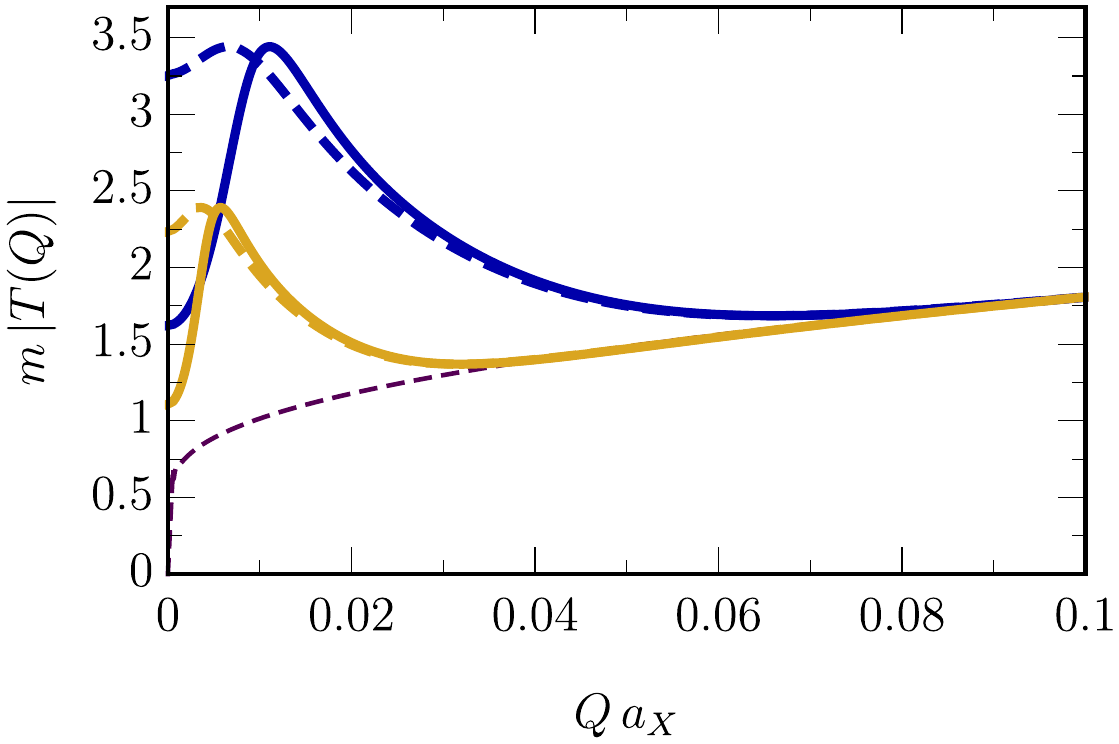}
	\caption{Polariton-electron scattering $T$ matrix at finite momentum. Blue (yellow) lines correspond to $T_{eP}(Q)$ with 
	$\Omega/\eb = 0.2$  $(0.05)$, where the solid and dashed lines are $\delta = -\Omega$ and $\Omega$, respectively, such that the Hopfield coefficients are the same for the two negative (positive) detuning lines.
		The thin purple dashed line is the low-energy exciton-electron $T$ matrix $ T_{eX}(Q) =  \frac{3\pi/m}{-\ln\left[-3\epsilon_Q/2\varepsilon_{1}\right]}$. 
	}
	\label{fig:Tfinite}
\end{figure}

\paragraph{Resonant-like scattering at finite momentum.---} 
The enhancement of polariton-electron scattering is even more pronounced at finite momentum. 
Figure~\ref{fig:Tfinite} clearly shows how the strength of polariton-electron scattering, $|T_{eP}(Q)|$, can be significantly larger than that of exciton-electron scattering in the momentum range relevant to the polariton system. 
Most notably, we find a strong resonance-like feature at momenta comparable to the inflection point, where the character of the polariton quickly changes from being photon- to exciton-dominated. 
The height of the peak is determined by the size of the light-matter coupling $\Omega$, 
with the sharpest peaks occurring at negative detunings. 
This resonance feature arises from the 
competition between the increasing exciton Hopfield coefficient $|X_-(\Q)|^2$
\jfl{and the energy-dependence of the underlying off-shell exciton-electron interactions~\cite{ePlong}.}
This is qualitatively different from the commonly applied Born approximation which, \jfl{for momenta $Q\ll a_X^{-1}$,} 
is well-described by the monotonic function $(4\pi/m)|X_-(\Q)|^2$ \cite{ePlong}. Furthermore, the Born approximation predicts that our solid (dashed) lines in Fig.~\ref{fig:Tfinite} would lie on top of each other, in contrast to the results of our fully microscopic calculation which clearly depend on the light-matter interaction strength.
We stress that the enhanced polariton-electron interaction at finite momentum is also distinct from the so-called optical parametric oscillation condition (see, e.g., Ref.~\cite{Ciuti2003}), 
 since the present resonance-like feature occurs in scattering where the total momentum is zero and the magnitude of the relative momentum is unchanged.

\paragraph{Conclusions and outlook.---} We have presented an exact microscopic description of polariton-electron interactions for the simplified case of strongly screened interactions between electrons and holes. In particular, our analytic expression for the exciton-polariton propagator provides a basis for future few- and many-body calculations within this model.
Our key finding is that strong light-matter coupling generically enhances polariton-electron interactions compared with the exciton-electron case, thus leading to  
a resonance-like peak in the polariton-electron scattering at finite momentum (Fig.~\ref{fig:Tfinite}).

Our results also apply to more realistic charge interactions and can be directly tested in doped semiconductor microcavities,
such as the scenario of Fermi polaron-polaritons~\cite{SidlerNatPhys16}. Here, the resonance-like feature should be observable in the transmission spectrum of a doped planar microcavity, where the in-plane momentum of the polariton can be varied (in contrast to the case in Ref.~\cite{SidlerNatPhys16}).
Moreover, unlike previous work, this interaction enhancement occurs in a spin-polarized system and 
does not rely on the presence of a trion bound state, thus allowing polaronic physics to be clearly separated from 
few-body bound states. This impacts the ongoing debate on whether one forms trions or 
polarons in a doped semiconductor~\cite{Efimkin2017,CombescotPRB2018,GlazovJCP2020}.
Finally, there is the prospect of using the sharp resonance-like peak to control quantum correlations between polaritons~\cite{Tan2020,kyriienko2019nonlinear}, since it can be used to enhance the effective polariton-polariton interactions induced by an electron medium. Alternatively, it could help achieve strong polariton-mediated interactions between electrons, ultimately leading to electron pairing and superconductivity~\cite{LaussyPRL10,CotlePRB16,KavokinNatMater2016}.


\acknowledgements
We thank Dmitry Efimkin, Eliezer Estrecho, Emma Laird, David Neilson, Elena Ostrovskaya, and Maciej Pieczarka for useful discussions, and we acknowledge support from the Australian Research Council Centre of Excellence in Future Low-Energy Electronics Technologies (CE170100039). JL is furthermore supported through the Australian Research Council Future Fellowship FT160100244.


\bibliography{e_P_scattering_refs}

\end{document}